 \documentclass[twocolumn]{aastex631}

\shorttitle{X-ray flash of YZ Ret}
\shortauthors{Kato et al.}

\begin{document}

\title{A light-curve analysis of the X-ray flash 
first observed in classical novae}



\author[0000-0002-8522-8033]{Mariko Kato}
\affil{Department of Astronomy, Keio University, 
Hiyoshi, Kouhoku-ku, Yokohama 223-8521, Japan} 
\email{mariko.kato@hc.st.keio.ac.jp}

\author{Hideyuki Saio}
\affil{Astronomical Institute, Graduate School of Science,
    Tohoku University, Sendai 980-8578, Japan}


\author[0000-0002-0884-7404]{Izumi Hachisu}
\affil{Department of Earth Science and Astronomy, 
College of Arts and Sciences, The University of Tokyo,
3-8-1 Komaba, Meguro-ku, Tokyo 153-8902, Japan} 



\begin{abstract}
An X-ray flash, expected in a very early phase of a nova outburst,
was at last detected with the {\it SRG}/eROSITA in the classical nova
YZ Reticuli 2020.  The observed flash timescale, luminosity, and blackbody
temperature substantially constrain the nova model. 
We present light curve models of the X-ray flash for various
white dwarf (WD) masses and mass accretion rates.
We have found the WD mass in YZ Ret to be as massive as
$M_{\rm WD}\sim 1.3 ~M_\sun$ with mass accretion rates 
of $\dot M_{\rm acc}\sim 5 \times 10^{-10}- 5\times 10^{-9}
~M_\sun$ yr$^{-1}$ including the case that the mass accretion rate 
is changing between them, 
to be consistent with the {\it SRG}/eROSITA observation. 
The X-ray observation confirms the luminosity to be close 
to the Eddington limit at the X-ray flash.
The occurrence of optically thick winds, with the photospheric 
radius exceeding $\sim 0.1~R_\sun$, terminated the X-ray flash 
of YZ Ret by strong absorption.
This sets an constrain on the starting time of wind mass loss.
A slight contamination of the core material into the hydrogen rich 
envelope seems to be preferred to explain a very
short duration of the X-ray flash.
\end{abstract}




\section{Introduction}
\label{introduction}

A nova is a thermonuclear runaway event on a mass-accreting white dwarf
(WD) \citep[see, e.g., ][for a recent self-consistent
calculation]{kat22sh}.
A hydrogen-rich envelope on the WD quickly brightens  
up to $L_{\rm ph} \sim 10^{38}$ erg s$^{-1}$ or several $10^4 ~L_\sun$
just after the start of a runaway of the hydrogen-shell 
burning.  Here, $L_{\rm ph}$ is
the photospheric luminosity.  In an early phase of expansion of the
photosphere, its surface temperature increases up to $T_{\rm ph} \sim
10^6$ K and the nova emits supersoft X-rays at a rate of $L_{\rm X}
\sim 10^{38}$ erg s$^{-1}$.  The duration of the bright soft X-ray phase
is so short that it is called the ``X-ray flash''
in the rising phase of a nova. Such X-ray flashes have long been
expected to be observed, but no one detected until
UT 2020 July 7 when the eROSITA instrument on board 
{\it Spectrum-Roentgen-Gamma (SRG)} scanned the region of YZ Reticuli
\citep{kon22wa}.

The nova outburst of YZ Ret (Nova Ret 2020) was reported 
first by \citet{mcn20} at visual magnitude 5.3 
on UT 2020 July 15.  This object was known 
as a cataclysmic variable (MGAB-V207), a novalike VY Scl-type 
variable with irregular variations 
in the $V$ magnitude range $15.8-18.0$ mag 
\citep{kil15ow}. The nova was classified as a He/N-type by \citet{car20sd}. 
The distance to the nova is estimated to be $d=2.53_{-0.26}^{+0.52}$ kpc
by \citet{bai21rf} based on the {\it Gaia}/eDR3 data.  The galactic
coordinates are $(\ell, b)= (265\fdg 3975, -46\fdg 3954)$ (ep=J2000),
so the nova is located at 1.8 kpc below the galactic disk.
The galactic absorption toward YZ Ret
is as low as $E(B-V) \sim 0.03$ \citep{sok22ll}.
The orbital period  was obtained by \citet{schaefer22} to be
$P_{\rm orb}=0.1324539 \pm 0.0000098$ days (=3.17889 hr).
Thanks to the short distance $d=2.5$ kpc from the Earth 
and very low galactic absorption
$E(B-V) \sim 0.03$, the nova was observed in mutiwavelengths,
including optical, X-ray, and gamma-ray from a very early phase of
the outburst until a very later phase through a supersoft X-ray source
(SSS) phase.  X-ray and $\gamma$-ray observations are reported by 
\citet{sok22ll} and \citet{kon22wa}.

The most remarkable characteristic of YZ Ret observation is 
the detection of an X-ray flash.
This is the first positive detection among any types of nova outbursts. 
\citet{kon22wa} reported the X-ray flash on UT 2020 July 7 observed 
with the {\it SRG}/eROSITA.  This detection is serendipitous during
its all sky survey.
Since the X-ray flash of a nova is a brief phenomenon that occurs 
 before the optical brightening, it is not 
possible to exactly predict when it occurs. 

Historically, there were a few attempts for detecting an X-ray flash.
\citet{mor16} searched MAXI data for X-ray flashes
that would possibly occurred during the MAXI survey
at the position and time of known nova outbursts, but unsuccessful. 
\citet{kat16xflash} attempted to detect an X-ray flash just before 
the expected nova outbursts of one-year-period recurrent nova 
M31N 2008-12a. This was the first planned observation, 
but no X-ray flash was detected in its 2015 outburst. 

Theoretical models predict X-ray flashes detectable only for a very short time
($\lesssim 1$ day) depending on the WD mass and mass-accretion rate
\citep{kat16xflash}. In low mass WDs, 
the surface temperature does not rise as high as to emit much 
X-rays, and most of photon energy is far-UV instead of X-ray
\citep[e.g,][]{kat17,kat22sh}. 
Thus, the X-ray flash should be detectable only in massive WDs. 
The time interval between the X-ray flash and optical maximum also
depends on the WD mass and mass-accretion rate,
which is, however, poorly understood. 
A very early phase, before a nova brightens optically,
is one of the frontiers in nova studies.
Because no planned observations had been successful, 
only the serendipitous detection of the X-ray flash with the {\it SRG}/eROSITA
gives us invaluable information on the very early phase of a nova.  

In this paper, we present theoretical light curve models
of X-ray flashes, the duration of which are short enough to match
the {\it SRG}/eROSITA observation.  Only massive WDs are responsible
for the flash like in YZ Ret. 

This paper is organized as follows.
Section \ref{section_model} presents our numerical method
and results for the X-ray flash, and compares them with
the observational data for YZ Ret.
Conclusions follow in section \ref{conclusions}.

\section{Model calculation of X-ray flash}
\label{section_model}

We have calculated models of nova outbursts with 
a Henyey type time-dependent
code combined with steady-state optically thick wind solutions.
The numerical method is the same as that in \citet{kat22sh}. 
We list our model parameters in Table \ref{table_models}. 
From left to right, model name, WD mass, mass accretion rate, additional 
carbon mixture in the hydrogen-rich envelope, recurrence period of 
nova outbursts, starting time of winds since the onset of 
thermonuclear runaway, maximum nuclear burning rate 
that represents the strength of a flash, ignition mass, and pass or not
the requirement from the scan detection.   
The 1.0 $M_\sun$ WD model (Model A) is taken from \citet{kat22sh}.
The mass accretion rate 
$\dot{M}_{\rm acc}= 5\times 10^{-9} ~M_\sun$ yr$^{-1}$
is close to the median value in the distribution of mass-accretion rates
for classical novae obtained by \citet{sel19g} while
the mass accretion rate 
$\dot{M}_{\rm acc}= 5\times 10^{-10} ~M_\sun$ yr$^{-1}$
is close to the empirical rate \citep{kni11bp}
for cataclysmic variable systems with an orbital period of 
$P_{\rm orb}= 3.18$ hr. 
Since many old novae are observed to fade significantly on 
timescales of $\sim$ 100 years \citep[e.g., ][]{due92, joh14}, 
we have taken into account a gradual decrease of accretion rate
in Model H, in which accretion resumes at a rate of 
$\dot M_{\rm acc}=5 \times 10^{-9}~M_\odot$yr$^{-1}$ just after 
the end of a previous flash, while the accretion rate gradually decreases. 

We have assumed solar composition for accreted matter 
($X=0.7$, $Y=0.28$, and $Z=0.02$) for all the models except Model J. 
In many classical novae, 
heavy element enrichment is observed in the ejecta \citep[e.g.,][]{geh98tw}.
To mimic such a heavy element enrichment 
one may replace the envelope composition with that polluted 
by the WD core composition 
at the onset of thermonuclear runaway 
\citep[e.g.,][]{sta20}, or assume a CO enhancement 
in the accreting matter \citep[e.g.,][]{chen19}. 
In Model E, F, G, H, and I we have increased carbon mass fraction by 0.1 
and decreased helium mass fraction by the same amount at the 
onset of thermonuclear runaway.  In Model J,
we have assumed carbon rich mixture ($X=0.6$, $Y=0.28$ $X_{\rm C}=0.1$, 
and $Z=0.02$) for accreting matter.

\begin{deluxetable*}{cllllrlllllll}
\tabletypesize{\scriptsize}
\tablecaption{Nova Models
\label{table_models}}
\tablewidth{0pt}
\tablehead{
\colhead{Model} & &
\colhead{$M_{\rm WD}$} &
\colhead{$\dot M_{\rm acc}$} &
\colhead{C mix} &
\colhead{$t_{\rm rec}$} &
\colhead{$t_{\rm ML}$\tablenotemark{a}}&
\colhead{$L_{\rm nuc}^{\rm max}$} &
\colhead{$M_{\rm ig}$} &
\colhead{comment} \\
\colhead{}&  
\colhead{}&  
\colhead{ ($M_\odot$) } &
\colhead{($M_\odot$ yr$^{-1}$)   } &
\colhead{ }&
\colhead{ (yr)}&
\colhead{(hr)}&
\colhead{($ 10^8 L_\odot $) } &
\colhead{($10^{-5}M_\odot$)   } &
\colhead{pass or not\tablenotemark{b}}
}
\startdata
A\tablenotemark{c} & ... & 1.0 & $5\times 10^{-9}$ & no  & 5400 & 25   & 2.3  & 3.0 & Fig.3a,no \\
B & ... & 1.2 & $5\times 10^{-9}$ & no    & 1500 & 18 & 3.6  & 0.82 & Fig.3b,no \\
C & ... & 1.35 & $5\times 10^{-9}$ & no    & 220 & 6.1  & 3.3 & 0.13 & Fig.3c,no \\
D & ... & 1.35 & $5\times 10^{-10}$ & no    & 2900 & 4.0   & 8.6  & 0.16 & Fig.3d,pass \\
E & ...& 1.3 & $5\times 10^{-9}$ & yes\tablenotemark{d}  & 300 & 3.2  & 16  & 0.16 & Fig.3e,pass  \\
F & ...& 1.3 & $5\times 10^{-10}$ & yes\tablenotemark{d}  & 4600 & 1.5  & 74  & 0.25 & Fig.3f,pass \\
G & ...& 1.35 & $5\times 10^{-9}$ & yes\tablenotemark{d}  & 120 & 1.8  & 12 & 0.070 & Fig.3g,pass  \\
H & ...& 1.35 & Decreasing\tablenotemark{e} & yes\tablenotemark{d}  & 1700 & 1.1 & 63 & 0.10 & Fig.3h,pass \\
I & ...& 1.35 & $5\times 10^{-10}$ & yes\tablenotemark{d}  & 1900 & 1.5  & 60 & 0.10 & Fig.3i,pass \\
J & ...& 1.35 & $5\times 10^{-10}$ & yes\tablenotemark{f}  & 1200 & 2.9  & 3.6 & 0.066 & ...,pass
\enddata
\tablenotetext{a}{Starting time of optically thick winds 
since the $L_{\rm nuc}$ peak ($t=0$).}
\tablenotetext{b}{pass or not the detection requirement of 22nd (no),
23rd (yes), and 24th (no) scan.}
\tablenotetext{c}{Model taken from \citet{kat22sh}.}
\tablenotetext{d}{Increased carbon mass-fraction by 0.1 at ignition.} 
\tablenotetext{e}{Mass accretion rate is changing from $5\times 10^{-9}$
to $5\times 10^{-10} ~M_\sun$ yr$^{-1}$.} 
\tablenotetext{f}{Mass accretion of carbon rich matter by 0.1.} 
\end{deluxetable*}

\subsection{Cycle of nova evolution in the HR diagram}
\label{outburst_model_hr_diagram}


\begin{figure*}
\epsscale{0.8}
\plotone{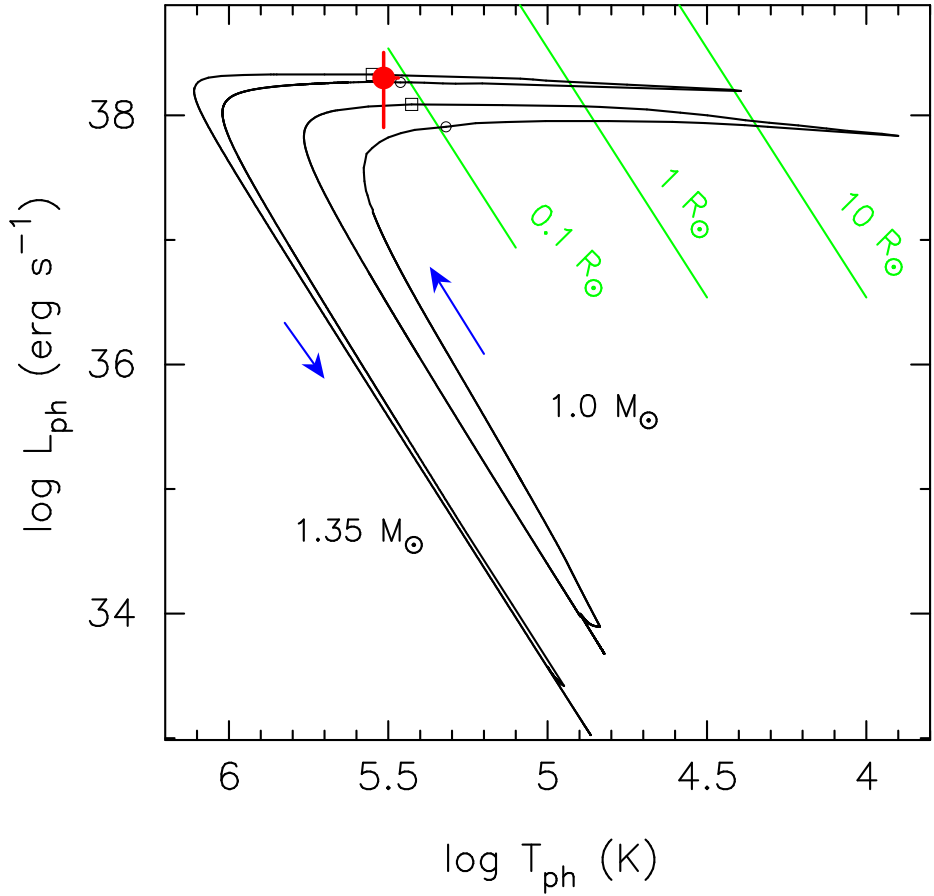}
\caption{
Comparison of nova outbursts in the HR diagram
between a 1.0 $M_\sun$ white dwarf (WD) (Model A)
and 1.35 $M_\sun$ WD (Model I), where $L_{\rm ph}$ and $T_{\rm ph}$ are
the photospheric luminosity and temperature, respectively.
The green lines show equi-radius lines, the value of which is attached
beside each line, based on the relation of $L_{\rm ph}=4 \pi R_{\rm ph}^2
\sigma T_{\rm ph}^4$, where $R_{\rm ph}$ and $\sigma$ are the 
photospheric radius and Stefan-Boltzmann constant, respectively.
Each arrow indicates the direction of evolution.   
The filled red circle with error bars corresponds to the X-ray flash
in YZ Ret  
\citep{kon22wa}.
The open circles indicate the start of optically thick winds
while the open squares represent the end of winds.
\label{hrcompari}
} 
\end{figure*}

Figure \ref{hrcompari} shows one cycle of nova outbursts for a 
1.0 $M_\sun$ WD (Model A) 
and 1.35 $M_\sun$ WD (Model I) in the HR diagram.
In the quiescent phase (inter outburst period), the accreting WD stays
around the bottom of each loop. 
After the thermonuclear runaway sets in, the WD goes upward 
keeping the photospheric radius almost constant.
The photospheric temperature increases to maximum, 
$\log T_{\rm ph}^{\rm max}$ (K)=5.58 in the 1.0 $M_\odot$ model 
(Model A) and 6.02 in the 1.35 $M_\odot$ model (Model I). 
In these high temperature phases, the WD photosphere emits X-ray/UV
photons, which corresponds to an X-ray/UV flash.
After that, the envelope expands and the photospheric temperature 
begins to decrease. 
Optically thick winds start when the envelope expands and the surface
temperature decreases to $\log T_{\rm ph}$ (K)=5.32 
in the 1.0 $M_\odot$ model and 5.46 in the 1.35 $M_\odot$ model 
(at each open circle). 

The filled red circle with error bars indicates the 
position of the X-ray flash of YZ Ret observed 
by {\it SRG}/eROSITA. The point lies on the evolution 
track of Model I (1.35 $M_\sun$) just before optically thick 
winds start, which is important because the winds possibly 
self-absorb soft X-rays as discussed in the next subsection.
  
When the photospheric radius attains its maximum expansion,
the wind mass loss rate also reaches maximum. 
The hydrogen-rich envelope mass quickly decreases
mainly due to wind mass loss. 
The photospheric radius begins to shrink while the photospheric temperature
turns to increase. In a later phase, optically thick winds stop
(at each open square) and the photospheric temperature 
becomes as high as $\log T_{\rm ph}$ (K) =5.5 - 6.1 and the WD
again emits X-ray/UV photons.  This phase is called the supersoft X-ray
source (SSS) phase in the decay phase of a nova outburst. 

The wind phase after the optical maximum and following SSS phase    
have been observed well in a number of nova outbursts. However, 
an early phase before the optical maximum has rarely been studied.


\begin{figure*}
\epsscale{0.8}
\plotone{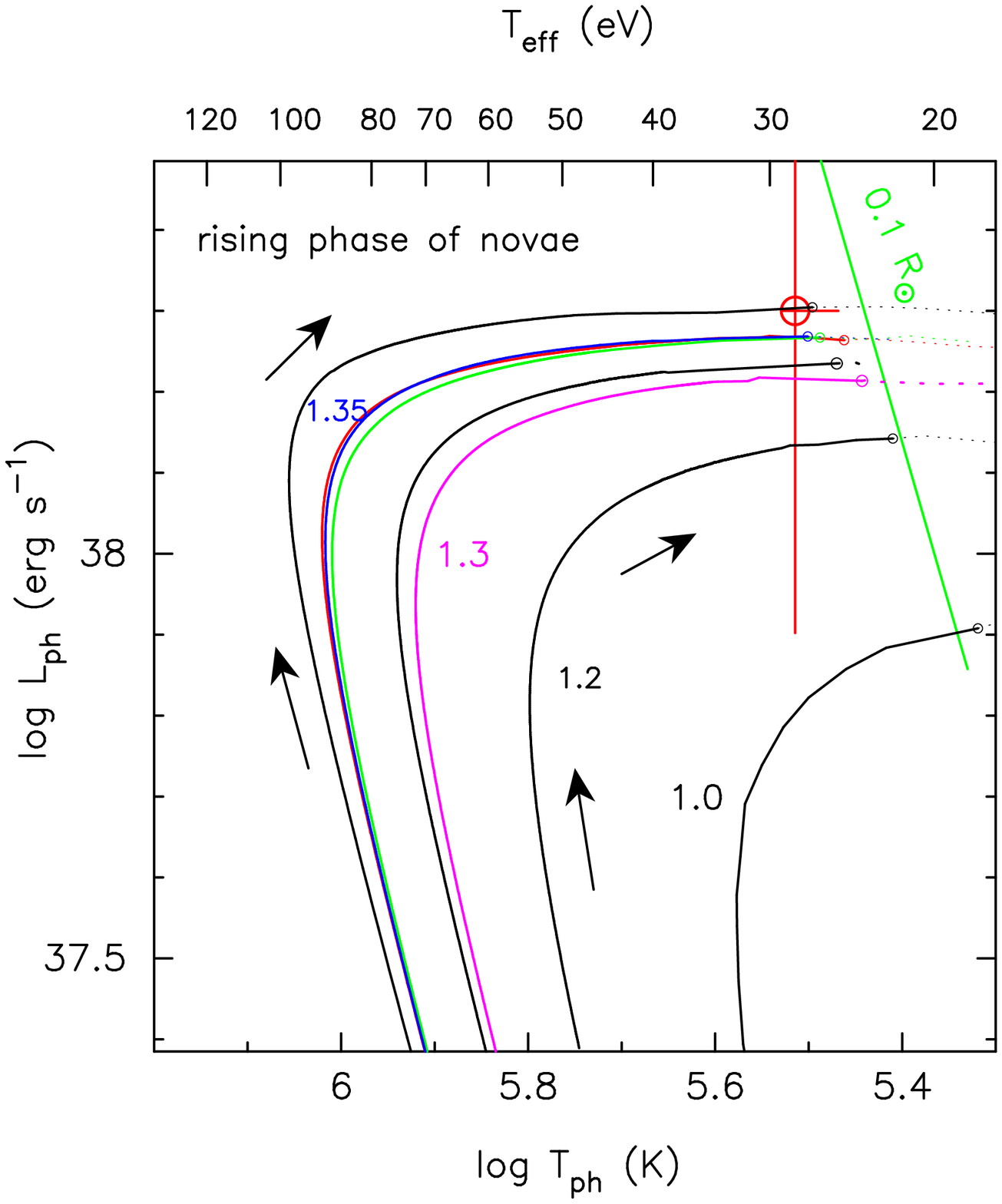}
\caption{
Close up view only of the rising phases of novae in the HR diagram. 
Optically thick winds start at the open circles on each track.
The track rightside the open circle (low temperature side) 
corresponds to the wind phase, denoted by a dotted line.
Each model corresponds to those in Table \ref{table_models},
from upper to lower, black line (Model J), red (I), blue (C),
and green (D), all for 1.35 $M_\odot$ WDs, 
Model E (black) and F (magenta) for 1.3 $M_\odot$ ,
B (1.2 $M_\odot$: black), and A (1.0 $M_\odot$: black).
The red dot with the error bars denotes 
k$T_{\rm ph}=28.2^{+0.9}_{-2.8}$ eV, 
$L_{\rm ph}=(2.0 \pm 1.2) \times 10^{38}$ erg s$^{-1}$, 
and radius $R_{\rm ph}= 50000\pm 18000$ km
(= $0.07~\pm 0.026~ R_\odot$) estimated for YZ Ret \citep{kon22wa}. 
\label{hr}}
\end{figure*}

\subsection{Optically thick winds absorb X-rays}
\label{rising_phase_before_winds}

Figure \ref{hr} shows a close up view of the rising phase on the HR diagram. 
The optically thick winds starts at the open circles on each line. 
The wind phase is denoted by the dotted line.
The optically thick winds would absorb and 
strongly weaken the X-ray flux. 
The optical depth for X-ray is estimated 
from equation (9) of \citet{li17mc} as
\begin{eqnarray}
\tau_X & \approx & 8\times 10^3 \left( {{{\dot M}_{\rm wind}} \over
{10^{-7} M_\sun ~ {\rm yr}^{-1}}} \right)
\left( {{r} \over {10^{10} ~{\rm cm}}} \right)^{-1} \cr
& & \times
\left( {{v_{\rm wind}} \over {400 ~{\rm km} ~ {\rm s}^{-1}}} \right)^{-1}
\left( {{E_X} \over {\rm keV}} \right)^{-2}.
\label{hard_xray_optical_depth}
\end{eqnarray}
Parameters of our models soon after the start of winds are, 
$R_{\rm ph} \sim 10^{10}$ cm and $\dot{M}_{\rm wind}
\sim 10^{-7}$ $M_\sun$ yr$^{-1}$.
Using these values in equation (\ref{hard_xray_optical_depth}), 
we find the optical depth of X-ray in the winds to be 
as high as $\tau_{\rm X}\sim 10^6$
for $E_X \sim 0.1$ keV. 
Thus, soft X-ray emission would be absorbed in the wind phase.
In other words, the X-ray flash could be terminated by 
the start of the winds. 

\citet{kon22wa} fitted the X-ray spectrum of YZ Ret with a blackbody
spectrum and obtained $T_{\rm BB}= 3.27^{+0.11}_{-0.33} \times 10^5$ K
(k$T_{\rm BB}= 28.2^{+0.9}_{-2.8}$ eV).   
They also derived absolute luminosity to be 
$L_{\rm ph}=(2.0 \pm 1.2) \times 10^{38}$ erg s$^{-1}$. 
These estimates are plotted in Figure \ref{hr} 
by an open red circle with error bars. 
The estimated blackbody flux has a large error bar, but it
is consistent only with relatively massive WDs 
($M_{\rm WD} \ga 1.2$ $M_\sun$).   

The estimated blackbody temperature is located leftside the open circles,
that is, before optically thick winds start. 
Thus, theoretically, no strong emission lines are expected.  
This is consistent with no prominent emission lines in 
the observed X-ray spectrum \citep{kon22wa}.

\citet{sok22ll} estimated the galactic hydrogen column density to be
$1\times 10^{19}$ cm$^{-2}$ $\lesssim N_{\rm H} 
\lesssim 1.86\times 10^{20}$ cm$^{-2}$.  \citet{kon22wa} obtained
$N_{\rm H} < 1.4\times 10^{20}$ cm$^{-2}$ toward the nova based on
their X-ray spectrum analysis and concluded
that there is no major intrinsic absorption during the X-ray flash.  
Such small hydrogen column density is consistent with our 
models in which the optically thick winds are absent 
at the stage of k$T_{\rm ph}=$ 28.2 eV.


\begin{center}
\begin{figure*}
\gridline{
          \fig{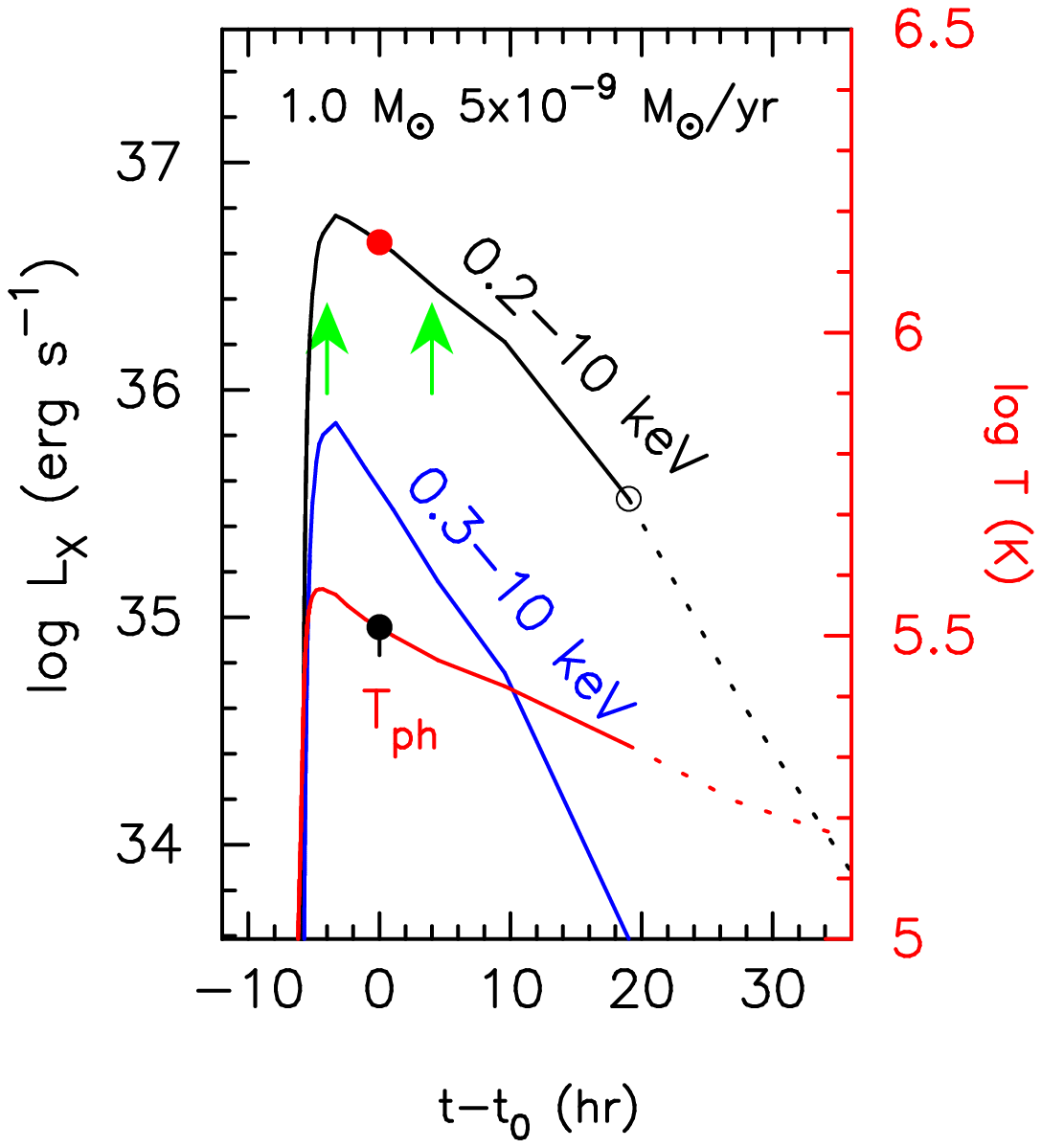}{0.28\textwidth}{(a)}
          \fig{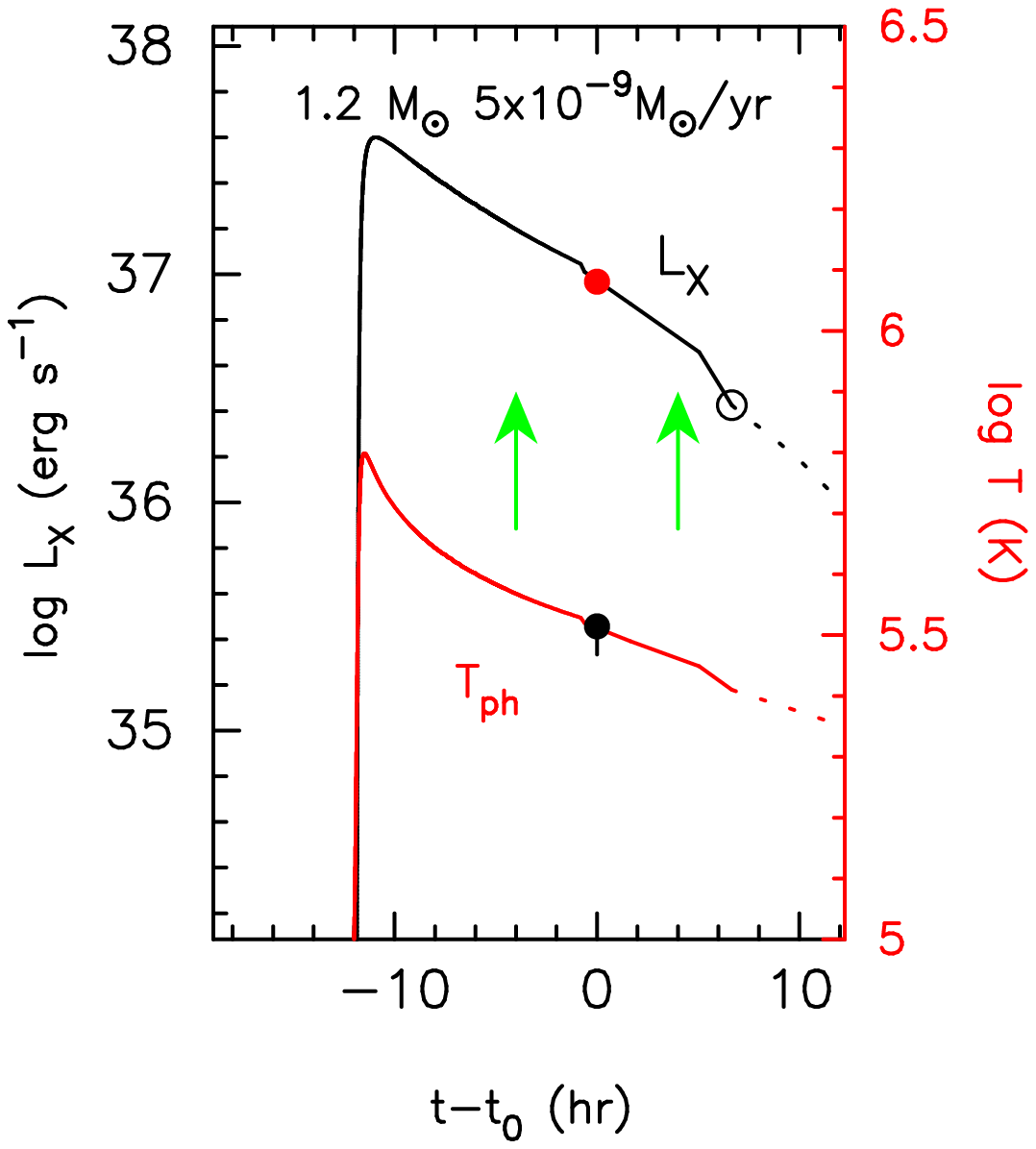}{0.28\textwidth}{(b)}
          \fig{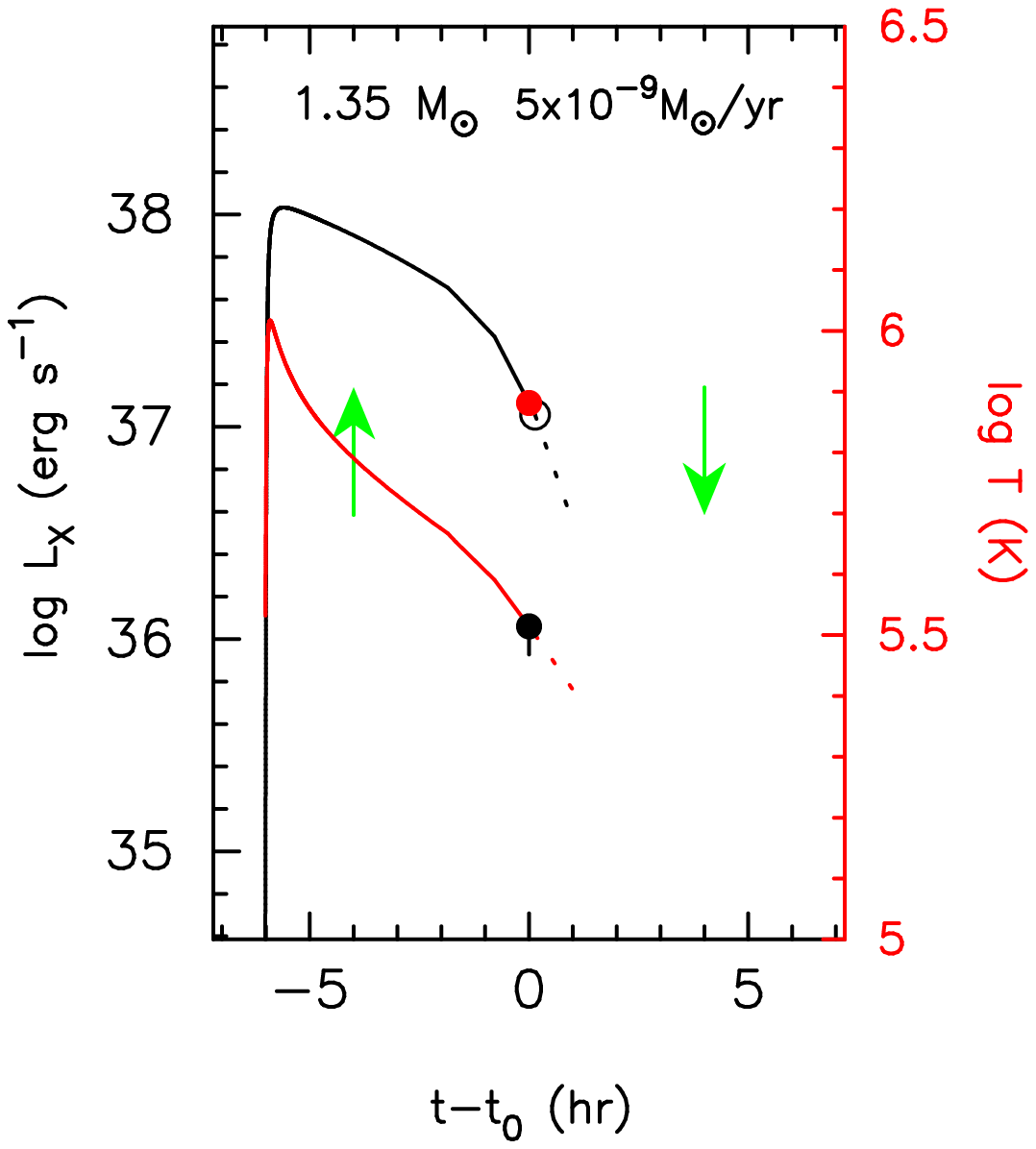}{0.28\textwidth}{(c)}
          }
\gridline{
          \fig{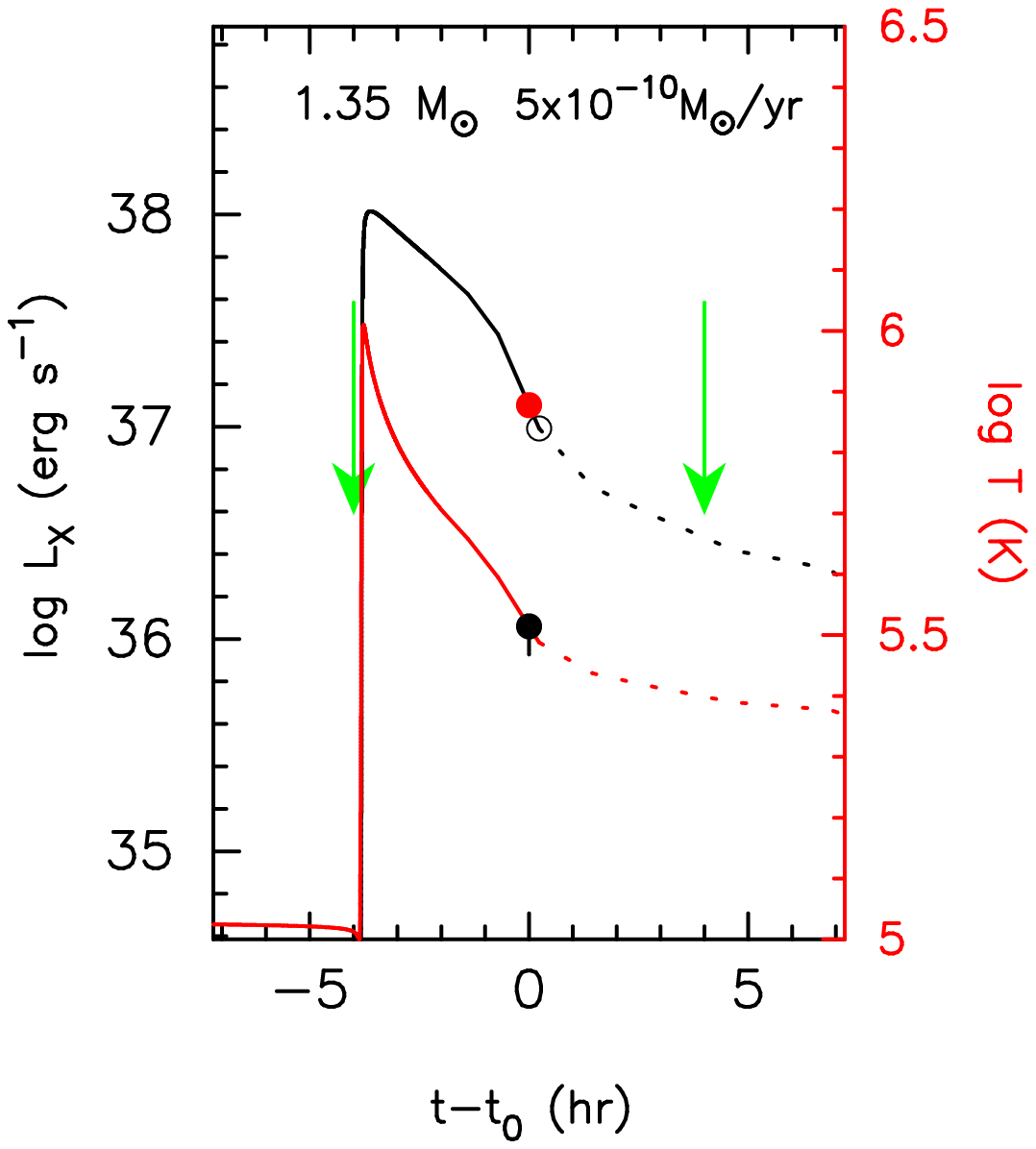}{0.28\textwidth}{(d)}
          \fig{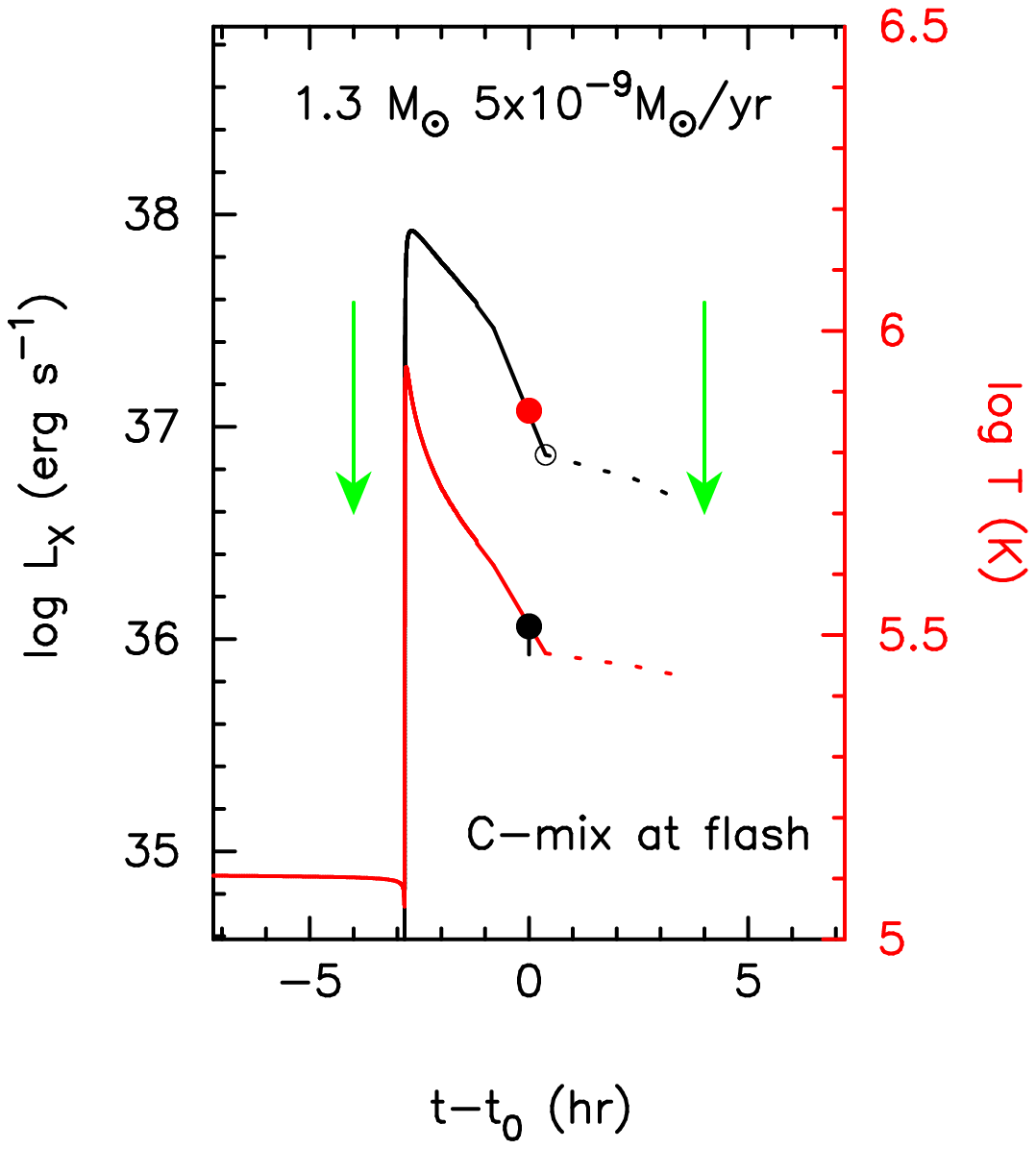}{0.28\textwidth}{(e)}
          \fig{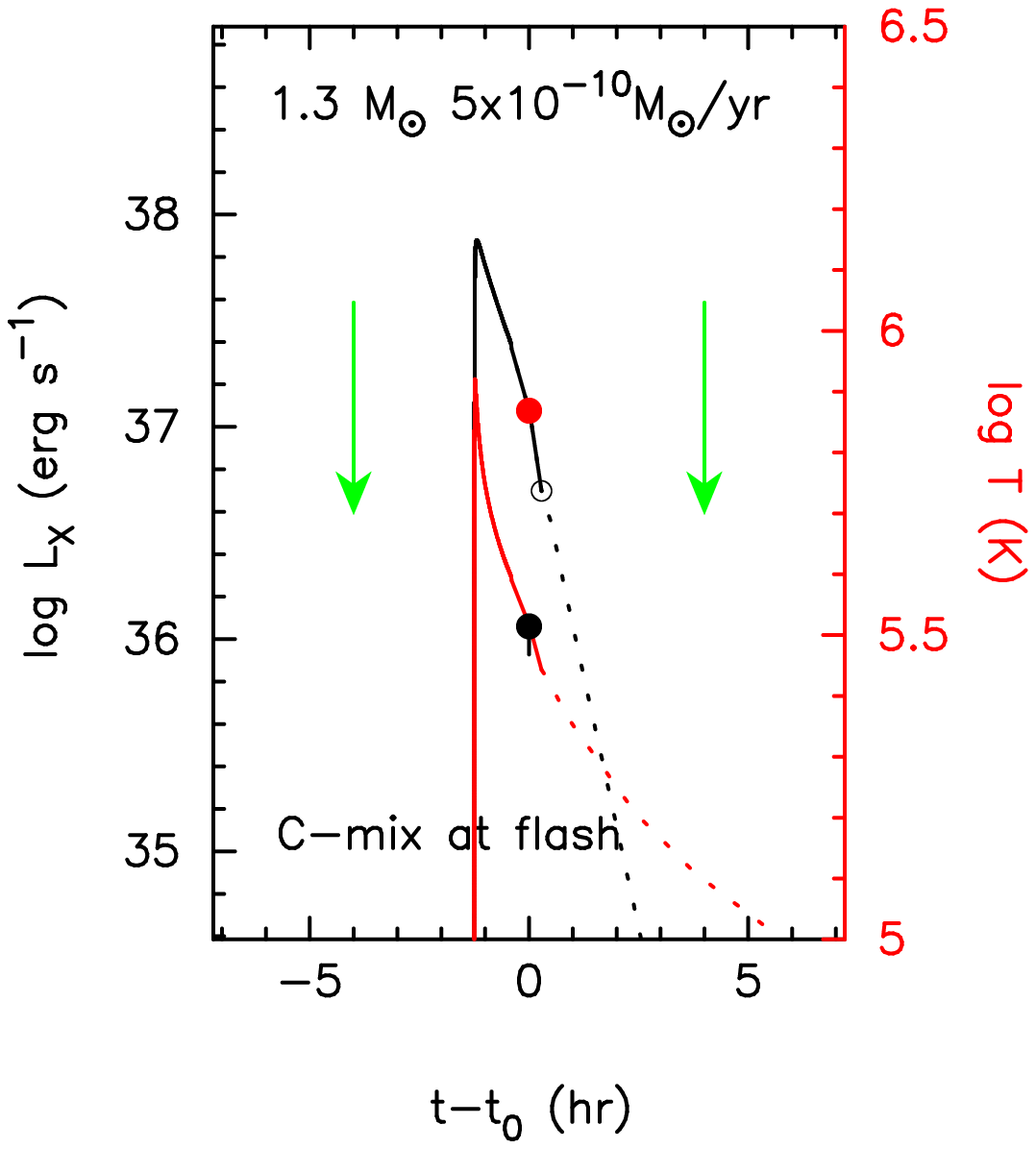}{0.28\textwidth}{(f)}
          }
\gridline{
          \fig{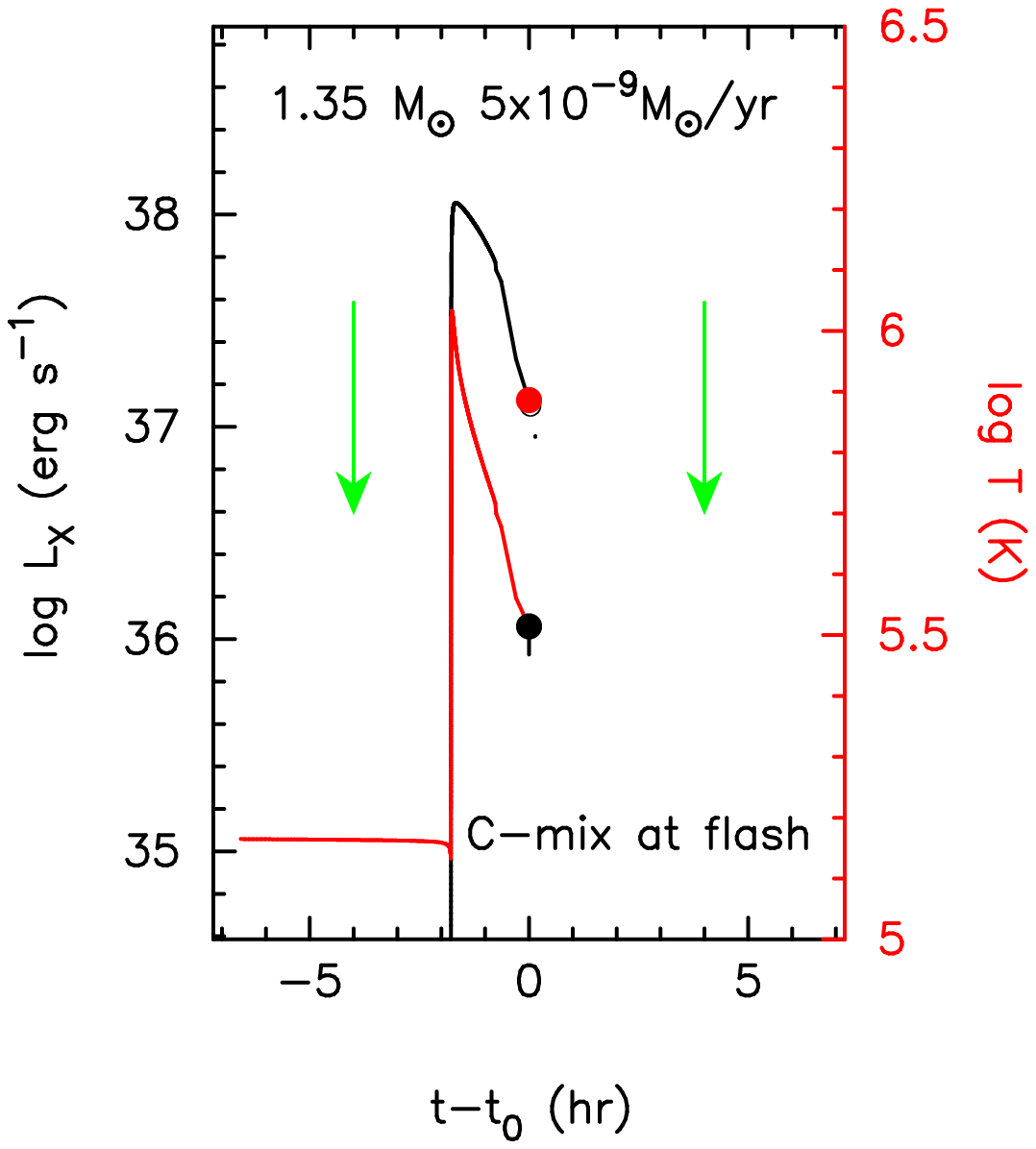}{0.28\textwidth}{(g)}
          \fig{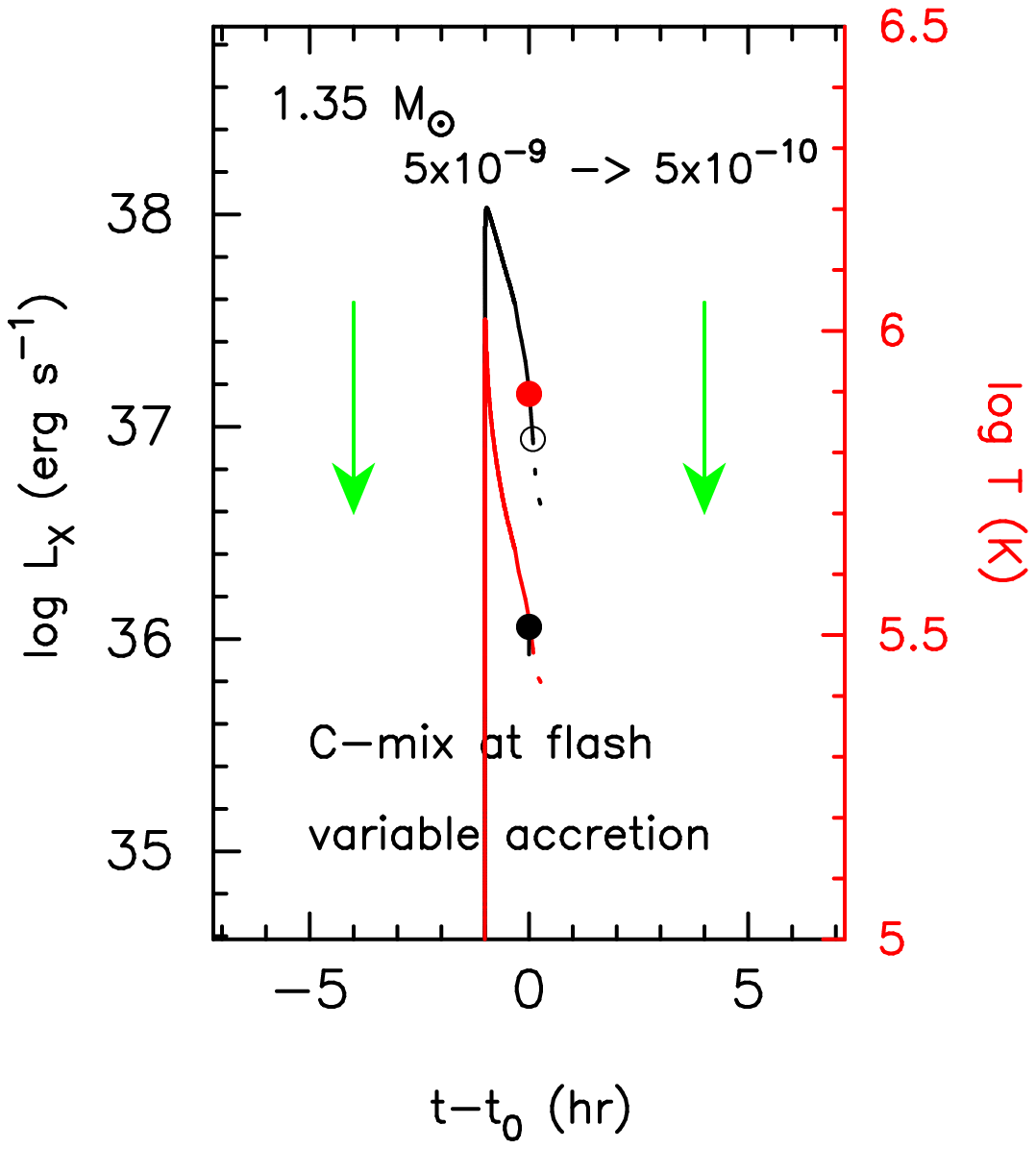}{0.28\textwidth}{(h)}
          \fig{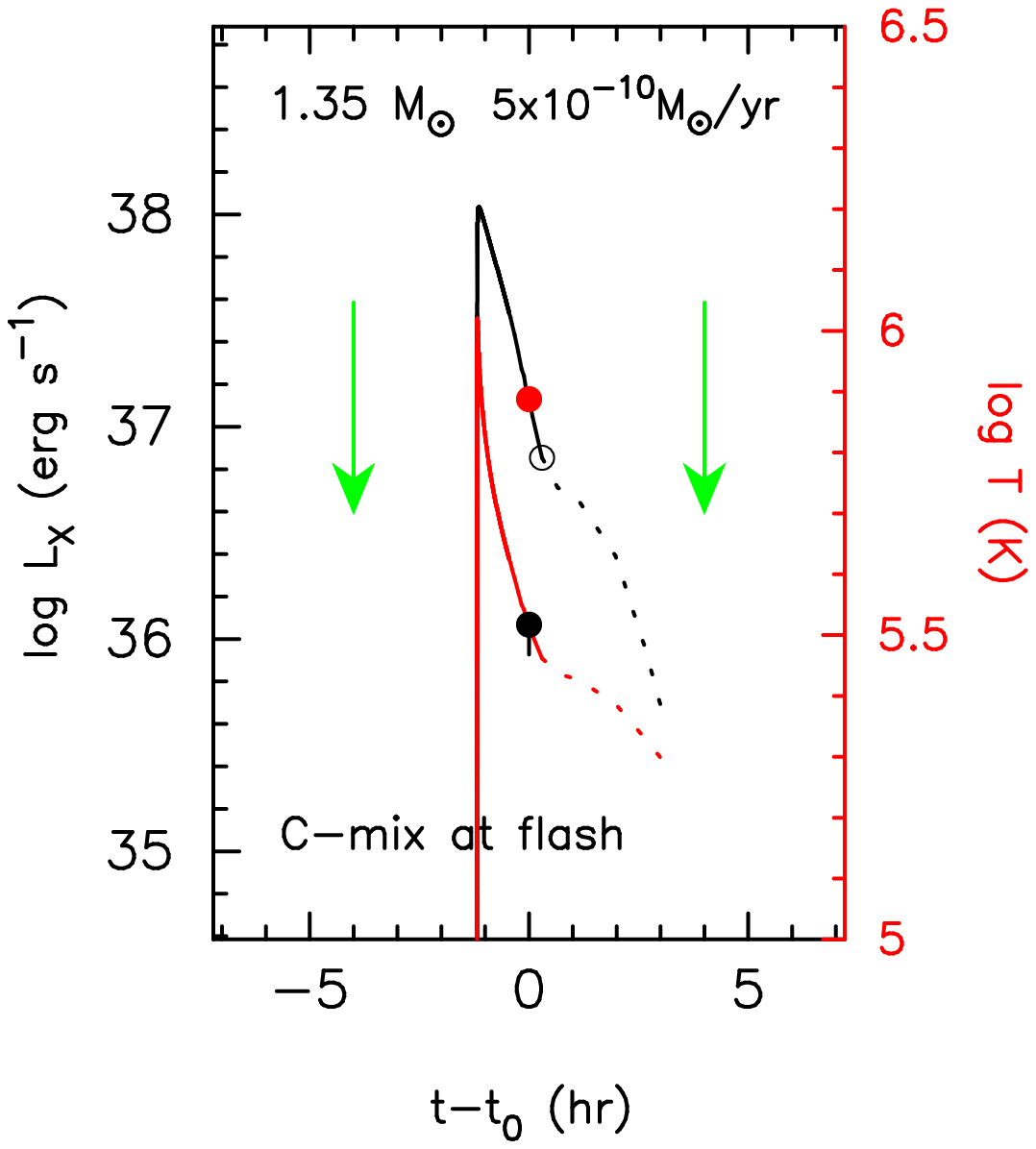}{0.28\textwidth}{(i)}
          }
\caption{
X-ray light curves for the X-ray flash models. 
Each panel corresponds to each model in Table \ref{table_models}.
The red lines are the temporal variations of the photospheric 
temperature, $T_{\rm ph}$, which are scaled on the right side 
vertical axis printed in red.  
The black lines show the estimated 0.2-10 keV ({\it SRG}/eROSITA)
band flux while the blue line in panel (a) indicates the (0.3-10) keV
({\it Swift}/XRT) band flux. They are scaled on the left vertical axis.
Each open circle on the black lines corresponds to 
the epoch when winds start.  The dotted part means a wind phase.
The large red dots on each black line show the X-ray luminosity
at the epoch of k$T_{\rm ph} =$28.2 eV ($\log T_{\rm ph} ~({\rm K})
= 5.515$), corresponding to the observed blackbody temperature 
of the YZ Ret flash. 
The origin of time denoted by $t_0$ is the observed time with
the {\it SRG}/eROSITA at the 23rd scan. The two green arrows 
indicate 4 hours before (22nd scan) and after (24th scan) 
the observation (23rd scan).  
\label{flashtheory}}
\end{figure*}
\end{center}

\subsection{Very short duration of X-ray flash}
\label{short_duration_x-ray_flash}

\citet{kon22wa} reported that the {\it SRG}/eROSITA scanned the
region of YZ Ret 28 times (every 4 hours) and detected YZ Ret at
the 23rd scan for about 36 s but did not detect 4 hours
before and after this scan.  This means that the duration of
X-ray flash is shorter than 8 hours.

Figure \ref{flashtheory} shows the X-ray light curves during 
the flash for our models in Table \ref{table_models} except model J. 
The black lines correspond to the 0.2-10 keV band of 
the {\it SRG}/eROSITA instrument while the red lines indicate
the temporal variations of photospheric temperature. 
The open circle on each black line indicates the start of optically
thick winds.  
The dotted line extending from the open circle indicates 
the evolution path with the winds, during which no 
significant X-ray flux is expected. 
The large black dot with error bars on each red line
shows the epoch when the photospheric temperature decreases to
k$T_{\rm ph}= 28.2_{-2.8}^{+0.9}$ eV.  The large red dot on each black
line corresponds to the X-ray luminosity at this epoch.
Two vertical green arrows indicate 4 hours before/after this epoch.
The X-ray luminosity at the epochs of arrows should be 
smaller than that at the red dot by four orders of magnitude 
because the {\it SRG}/eROSITA did not detect X-rays.

Model A, B, and C evolve slowly, and should have detectable 
X-ray fluxes even before and/or after four hours of the 
detection epoch by the {\it SRG}/eROSITA, contradicting 
with the non-detection. The upward (downward) arrow indicates 
the theoretical prediction of detectable (non-detectable) X-ray flux. 
Model D is marginally consistent with the requirement.

The carbon mixture models evolve 
much faster and easily fulfill the requirements. 
The CNO enrichment in the hydrogen rich envelope make flash 
evolution faster because the CNO reaction rates increase. 
A faster evolution is favorable to be consistent with 
the constraints from the observation of 
{\it SRG}/eROSITA. 
Furthermore, the appearance of the optically thick winds would  
contribute to shorten the duration of the X-ray flash of YZ Ret. 

 In Model H the mass accretion restarts 
after a shell flash ends 
with a rate of $\dot{M}_{\rm acc}= 5\times 10^{-9}$ $M_\sun$ yr$^{-1}$ 
which gradually decreases to 
$ 5\times 10^{-10}$ $M_\sun$ yr$^{-1}$ in the first 100 years   
of the quiescent phase and keep constant after that. 
Until the next outburst, long after 1600 years, 
the WD thermal structure is adjusted 
to the lower accretion rate. As a result 
the outburst properties should be similar to a model of mass 
accretion rate of $ 5\times 10^{-10}$ $M_\sun$ yr$^{-1}$. 
Model H shows in fact similar properties to model I, but 
slightly stronger flash properties, i.e., shorter flash duration and 
larger $L_{\rm nuc}^{\rm max}$ than in Model I.

\citet{kon22wa} also reported that the X-ray flux of YZ Ret 
decreased by about 
a few to 10\% even during the very short 36 s observation period.
We estimate  X-ray decay rates of our models to find  
a few \% in Model F, G, H, and I during a 30 s period
near the point denoted by the large red dot, being broadly consistent
with the observation.  

YZ Ret is a novalike VY Scl-type star, in which dwarf nova outbursts are
suppressed.  To suppress thermal disk instability of a dwarf nova,
the mass transfer rate is higher than $\dot{M}_{\rm crit}$.
This critical rate is estimated to be
$\dot{M}_{\rm crit}\sim 2 \times 10^{-9} ~M_\sun$ yr$^{-1}$ 
for $P_{\rm orb}= 3.18$ hr and an assumed total binary mass of
$M_1+M_2 \sim 1.3+ 0.3 = 1.6 ~M_\sun$ 
\citep[see, e.g., equations (3) and (4) of][]{osa96}.
Our Model E and G (1.3 and 1.35 $M_\sun$ WD models
with a relatively high mass accretion rate of $\dot{M}_{\rm acc} 
= 5\times 10^{-9} ~M_\sun$ yr$^{-1}$) satisfy this requirement
of novalike stars, i.e., $\dot{M}_{\rm acc} > \dot{M}_{\rm crit}$.   

We should emphasize the importance of low energy sensitivity of detector. 
The blue line in 
Figure \ref{flashtheory}a shows an X-ray light curve of the 
0.3-10 keV band corresponding to the {\it Swift}/XRT. 
The flux is about ten times smaller than the {\it SRG}/eROSITA flux
(0.2-10 keV band), clearly showing that,
for an efficient detection of X-ray flashes,
the low energy sensitivity (down to 0.2 keV or lower) is important.

\subsection{Clue to the origin of the super-Eddington luminosity}
\label{super_eddington}
We should remark the importance of the X-ray flash on the super-Eddington
problem in novae.
The bolometric luminosity of a star in hydro-static balance cannot
exceed the Eddington limit as long as spherical symmetry is assumed. 
The Eddington limit is defined by
\begin{eqnarray}
L_{\rm Edd} & \equiv & {{4\pi c G M_{\rm WD}} \over {\kappa}} \cr
 & = & 2 \times 10^{38} {\rm erg~s}^{-1}  
\left( {{1.7} \over {1+X}} \right) 
\left( {{M_{\rm WD}} \over {1.4 ~M_\sun}} \right)
\end{eqnarray}
for massive WDs with the mass of $M_{\rm WD}$, 
where $\kappa = 0.2 (1+X)$ cm$^2$ g$^{-1}$ is the electron scattering opacity.

YZ Ret reached its optical peak $V=3.7$ 
four days after the X-ray flash \citep{mcn20ph}.  
This brightness corresponds to an absolute $V$ magnitude 
of $M_V= 3.7 - (m-M)_V= -8.4$, which is several times larger than
the Eddington limit.\footnote{The bolometric magnitude
is $M_{\rm bol}= -7.07$ for $L_{\rm ph}= 2\times 10^{38}$ erg s$^{-1}$.
This corresponds to $M_V\approx -7.0$ or $-6.0$ for the bolometric
correction B.C.$=-0.07$ (8700 K) or B.C.$=-1.07$ (14000 K), for example.
Then the flux ratio is $10^{(8.4-7.0)/2.5} \approx 4$ or 
$10^{(8.4-6.0)/2.5} \approx 9$ (times the Eddington limit).} 
Here, the distance modulus in $V$ band is estimated to be
$(m-M)_V= 5 \log (d/10{\rm ~pc}) + A_V= 12.0+0.1=12.1$.

At the X-ray flash of YZ Ret, however, the total photospheric luminosity 
was estimated to be $L_{\rm ph}= (2.0\pm 1.2)\times 10^{38}$ erg s$^{-1}$
\citep{kon22wa}. 
Thus, the photospheric luminosity did not
largely exceed the Eddington limit  
(see Figures \ref{hrcompari} and \ref{hr}).
This clearly confirms that the nova envelope is in hydro-static
balance at the X-ray flash, 
when the optically thick winds had not yet started. 
Thus, we may conclude that the origin of super-Eddington
luminosity is closely related to the occurrence of optically thick winds.

\section{Conclusions}
\label{conclusions}

An X-ray flash in the rising phase of a nova was first detected 
in the classical nova YZ Ret, which provides us with invaluable information
for the nova physics. We may conclude our theoretical analysis as follows. 

\begin{itemize}
\item \citet{kon22wa} found X-ray spectrum at the flash to be consistent 
   with an unabsorbed blackbody of $3\times 10^5$ K. 
   This is consistent with our hydrostatic evolution models 
just before optically thick winds start. 
\item The blackbody temperature $T_{\rm BB}\approx
   3 \times 10^5$ K and luminosity of $L_{\rm ph} \approx 2\times
   10^{38}$ erg s$^{-1}$ are consistent with our models of very 
   massive WDs ($M_{\rm WD} \gtrsim 1.2 ~M_\sun$).
\item The very short duration of the X-ray flash ($\lesssim 8$ hr) further 
   constrains the mass of WD ($M_{\rm WD} \gtrsim 1.3~M_\odot$),
   depending on the degree of WD core material mixing into the
   hydrogen-rich envelope.  
   The generation of optically thick winds when the 
   photospheric radius exceeds $\sim 0.1~R_\sun$ might terminate 
   the X-ray flash of YZ Ret. 
\item YZ Ret is a novalike VY Scl-type star, which requires a relatively
   high mass accretion rate ($\ga \dot{M}_{\rm crit}\sim (2-3)
   \times 10^{-9} ~M_\sun$ yr$^{-1}$)
   to suppress dwarf nova outbursts.  If it is the case,
   our 1.3 and 1.35 $M_\sun$ WD models with a relatively high mass
   accretion rate of $\dot{M}_{\rm acc} = 5\times 10^{-9} ~M_\sun$ yr$^{-1}$
   (Model E and G) satisfy the requirement of 
   $\dot{M}_{\rm acc} > \dot{M}_{\rm crit}$.   
\item The nova envelope is in hydro-static balance at the X-ray flash,
   just before optically thick winds start.  
   In a few days later, the optical luminosity highly exceeds the Eddington
   limit. 
   This suggests that the origin of super-Eddington
   luminosity is closely related to the occurrence of optically thick winds. 
\end{itemize}



\begin{acknowledgments}
 We are grateful to the anonymous referee for useful comments,
 which improved the manuscript.
\end{acknowledgments}

\clearpage







\end{document}